\documentstyle[12pt]{article}

\begin{document}

\def \eeww{e^+e^- \to W^+W^-}
\def \em {\epsilon_1}
\def \ep {\epsilon_2}

\begin{center}
{\LARGE Effects of TeV Scale Gravity on $e^+e^-\to W^+ W^-$}
\\[8mm]
{\large P. Poulose\footnote{poulos@physik.rwth-aachen.de}
}\\[3mm]
{\it Institute of Theoretical Physics E, RWTH Aachen,}\\
{\it D-52056 Aachen,Germany}\\[2mm]
\end{center}

\begin{abstract}
We study the process $e^+e^-\to W^+W^-$  to probe low-scale gravity at
high energy linear colliders.  A characteristic signature of
extra-dimensions models is the forward-backward asymmetry due to the
interference of spin-2 graviton and spin-1 SM gauge boson exchange terms,
even when right-polarised electron beam is used.
Our analysis shows that larger than 5\% asymmetry is possible at a linear
collider with $\sqrt{s}=500$ (800) GeV if the mass scale, $M_S$ is smaller
than 2.7 (4.5) TeV.
$W^-$ polarisation factors measured with a few percent accuracy will
also be able to put similar limits on $M_S$.
\end{abstract}

\section{Introduction}

In this paper we study the process $\eeww$ in the presence of low-scale 
gravity with extra spatial dimensions.  In such a large extra-dimensions
scenario gravity is allowed to propagate in $D=4+n$ dimensions, while SM
particles confine to a sub-space of 4-dimensions.  This idea was proposed
by Arkani-Hamed, Dimopoulos and Dvali \cite{add}, and the phenomenological
consequences at colliders have been looked at in various processes in detail 
by many authors \cite{reviewXD}.  The main idea is to have a universally 
fundamental-scale, $M_S$, of TeV range where all interactions including 
gravity are comparable in strength. This is
contrary to the traditional belief that gravity becomes comparable in
stregth to the other interactions only at Planck mass, $M_P\sim 10 ^{19}$ GeV.
This lowering of the scale is possible if we consider $n$ spatial
dimensions in addition to the usual 3+1 space-time manifold. These
additional dimensions are compactified to a radius $R$.  Relation between 
the two mass scales, the size of the extra dimensions and the number of
extra dimensions is obtained by demanding that for distances larger than
$R$ gravitational potential behaves like the Newtonian $1/r$. Thus we get
the relation\cite{add}
\[M_P^2 = 8\pi R^n M_S^{n+2}.\]

With an $M_S\sim$ TeV, $n=1$ requires $R$ to be of the order of $10^{11}$
meters.  This is ruled out, since Newtonian gravity is tested to be 
correct up to about millimeter
level. $n=2$ or larger is allowed by this criterion. Thus gravity 
can propagate in $4+n$ dimensions as long as the extra $n$ dimensions are 
smaller than 1 mm. Since Standard Model (SM) interactions are tested up to
sub-fermi level, we require to confine these particles to a 4 dimensional
sub-space.\footnote{Alternative scenarios with SM particles allowed to 
move in the bulk are discussed in the literature. There is also the
scenario proposed by Randall and Sundrum\cite{RS} with non-factorizable
metric. We will not consider any of these cases here.} Massless gravitons
propagating in the bulk is seen from the 4 dimensional sub-manifold as
massive Kaluza-Klein (KK) modes with spin-0, spin-1 and spin-2.  Mass spectrum
can be treated to be continuous owing to the fact that mass splitting is
of the order of $1/R$, which is about $10^{-4}$eV in the case of $n=2$, 
and of the order of MeV in the case of $n=6$. 
These KK modes couple to matter and gauge particles through the 
energy-momentum tensor and its trace.  The new interactions influence
collider experiments in two ways.  One through the production of real KK
modes, and two through the exchange of virtual KK modes in various
processes.
For a review on the effects of this idea in linear colliders like TESLA,
see \cite{TESLA}.

In the following we will consider the process $\eeww$ to study the effect
of low-scale gravity at the proposed linear colliders to run at 500 GeV 
and above. 
Earlier studies of this process \cite{agashe,balazs} have discussed 
possible limits on $M_S$ obtainable measuring, mainly, deviation of the
total cross section from its SM value. 
We will discuss, in addition to the total cross section, 
other observables like the forward-backward asymmetry and
polarisation fraction of the $W$'s. 
In a linear collider it is possible to use polarised beams. We will discuss
the benifits of beam polarisation in the present case.

In section 2 we discuss the process and give the expressions for matrix 
elements. Section 3 describes different observables that are relevant 
at a high energy linear collider. Conclusions will be given in section 4.

\section{The process}

The process $\eeww$ has the standard interactions with the exchange of
photon and $Z$-boson in the $s$-channel and a $t$-channel with $\nu$
exchange. In the case of large extra dimensions, there are 
additional $s$-channels with spin-2 KK excitations of bulk gravitons. 
Spin-1 modes do not
couple to the SM particles, and spin-0 contributions are
proportional to the mass of the fermion invovled. We treat electron as
massless, and so neglect spin-0 exchange contribution. 

\begin{figure}[ht]
\vskip 7cm
\includegraphics{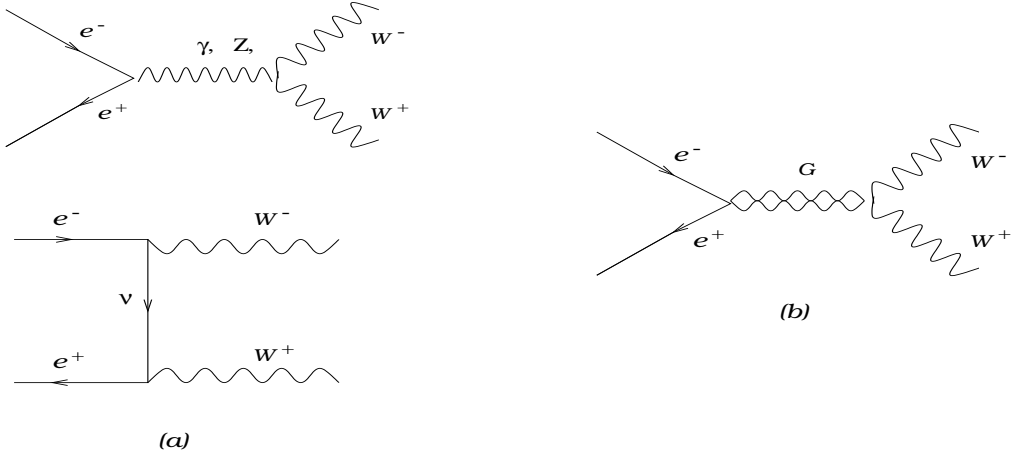}
\caption{Feynman diagramms of the process $\eeww$. (a) SM channels and
(b) Spin-2 KK exchange channel.}
\label{fig_fdiag}
\end{figure}

The Feynman diagrams of different channels are given in Fig. \ref{fig_fdiag}. 

Relevant Feynman
rules are derived by Han, Lykken and Zhang \cite{HanLykken}, and also
by Guidice, {\it et al} \cite{Guidice} in a linearised-gravity
aproximation.
There is an infinite tower of massive KK excitations all having a universal 
coupling to the SM particles.  Han, Lykken and Zhang \cite{HanLykken} have
derived the propagator factor after integrating over the spectrum, cutting 
it at the scale $M_S$.
Following them we get KK exchange contribution to the matrix element 
given by

\begin{eqnarray}
{\cal M}_G&=-2\pi C_4&\left\{\left[2 (k_1.q_2-k_1.q_1) (\em.\ep) +
2 (k_1.\em) (q_1.\ep) -2 (k_1.\ep) (q_2.\em) \right] 
(\bar v \slash \!\!\!q_1 u) \right.  \nonumber \\
&&+\left[-2 (k_1.q_2) (q_1.\ep)+s (k_1.\ep)\right] 
(\bar v \slash \!\!\!\em u) \nonumber \\
&&\left.+\left[-2 (k_1.q_1) (q_2.\em)+s (k_1.\em)\right] 
(\bar v \slash \!\!\!\ep u) \right\} 
\end{eqnarray}

Here $k_1$ is the momentum of initial $e^-$, and $u$ and $v$ are the
spinors corresponding to $e^-$ and $e^+$. $q_1$ and $q_2$ are
the momenta, and $\em$ and $\ep$ the polarisation vectors of $W^-$ and
$W^+$ respectively. $\sqrt{s}$ is the centre of masss energy of the
collider, and $C_4$ is the factor corresponding to KK propagators after 
integrating over all the modes.

The propagator factor in the limit of $M_S^2\gg s$ is given by
\begin{eqnarray}
C_4 &=& -\frac{{\rm log}(M_S^2/s)}{M_S^4}\;\;\;\;\;\;\;(n=2)\nonumber\\
	&=&-\frac{2}{(n-2)\;M_S^4}\;\;\;\;\;\;\;\;\;(n>2)
\end{eqnarray}

Here $n$ is the number of large extra dimensions.
In the case of high energy colliders expected to operate at energies 
of 500 GeV through 1-2 TeV, $M_S^2\gg s$ may not be a good approximation. 
In that case, nonresonant part of the propagator factors, neglecting 
the narrow width of each KK mode, as given by \cite{HanLykken} are,

\begin{eqnarray}
C_4 &=& \left(\frac{s}{M_S^2}\right)^{n/2-1}\;\frac{2\;I_n}{M_S^4} \\
I_n &=&\left\{\begin{array}[c]{l}
	 -\sum_{k=1}^{n/2-1} \frac{1}{2k}\left(\frac{M_S}{\sqrt{s}}\right)^{2k}
	-\frac{1}{2}{\rm log}\left(\frac{M_S^2}{s}-1\right)\;\;\;\;\;\;
	\;\;\;\;\;\;\;({\rm even}\;\; n)\\
 -\sum_{k=1}^{(n-1)/2} \frac{1}{2k-1}\left(\frac{M_S}{\sqrt{s}}\right)^{2k-1}
	+\frac{1}{2}{\rm log}\left(\frac{M_S+\sqrt{s}}{M_S-\sqrt{s}}\right)
	\;\;\;\;\;\; ({\rm odd}\;\; n)
	\end{array}
	\right.\nonumber 
\end{eqnarray}

Apart from these non-resonant contributions, there is also a resonant
contribution to the propagator factor, which adds an imaginary part to the
matrix element \cite{HanLykken}. 
In the case of $\eeww$, since the SM matrix elements are real, there is no
contribution from this part in the interference term. Purely gravity mediated
contribution will be of order $s^4/M_S^8$, and is negligible unless 
$\sqrt{s}$ is very close to $M_S$.

Explicit expressions for different helicity amplitudes are given in
\cite{agashe}. We consider full propagator factor including its dependence
on the number of extra dimensions, as given above. 
Thus to compare with the expressions given in \cite{agashe} the factor
$\lambda \frac{4}{M_S^4}$ should be replaced by $-2\pi C_4$.

In the next section we discuss various observables that could be used 
at a high energy linear collider.

\section{Observables}

Possible 
limits on $M_S$ that could be obtained at future linear colliders (LC) 
from considering the deviation of total cross section from the SM 
expectation were obtained in earlier studies \cite{agashe,balazs}.
Limits obtained depend on the aproximations used to obtain KK-propagator
factor.
Our main aim in this report is to identify possible signatures which 
characterize the spin-2 nature of the graviton exchange. The angular
distribution of the $W$'s and the forward-backward asymmetry, as we will see,
are good candidates in this respect. 

\subsection{Unpolarised $W$'s}

We first consider typical limit on $M_S$ that one could obtain at
an LC. 
In Table \ref{table-mslt} we give limits on $M_S$ that would correspond to
a 2\% deviation in the SM cross section at LC with c.m. energies of 500 GeV 
and 800 GeV.  We assume 100\% efficiency, and have not considered any
experimental cuts. More realisitic considerations will change the limits 
somewhat.
Large beam polarisations can be achieved in linear colliders.  Using a
right-polarised electron beam switches off the $\nu$ exchange contribution,
and hence improves the sensitivity to graviton exchange contribution.
At a 500 GeV collider, 
while using unpolarised beams or left-polarised electron beam with
unpolarised positron beam could reach $M_S$ values from about 1.3 TeV (for 
$n=2$) up to about 1.8 TeV (for $n=6$), right-polarised electron 
beam with unpolarised positron beam could reach up to about 2.7 TeV.

\begin{table}[ht]
\begin{center}
\begin{tabular}{c|c|c|c|c|c|c}
\hline
&&\multicolumn{5}{|c}{Limit on $M_S$ in TeV}\\ \cline{3-7}
$\sqrt{s}$&$e^-$ beam pol&$n=2$&$n=3$&$n=4$&$n=5$&$n=6$\\[1mm] \cline{1-7}
&&&&&\\
500 GeV&Unpol$\;/\;e^-_L$&1.8&1.7&1.5&1.4&1.3\\
&$e^-_R$&2.7&2.2&1.9&1.7&1.6\\[2mm] \cline{1-7}
&&&&&\\
800 GeV&Unpol$\;/\;e^-_L$&2.6&2.5&2.3&2.1&2.0\\
&$e^-_R$&4.5&3.6&3.2&2.9&2.7\\[2mm] \cline{1-7}
\end{tabular}
\caption{$M_S$ values corresponding to 2\% deviation in the 
SM cross section at LC with $\sqrt{s}=500$
GeV and $\sqrt{s}=800$ GeV for different beam polarisations. $n$ is the
number of extra dimensions.}
\label{table-mslt}
\end{center}
\end{table}

\begin{figure}[ht]
\vskip 5cm
\includegraphics{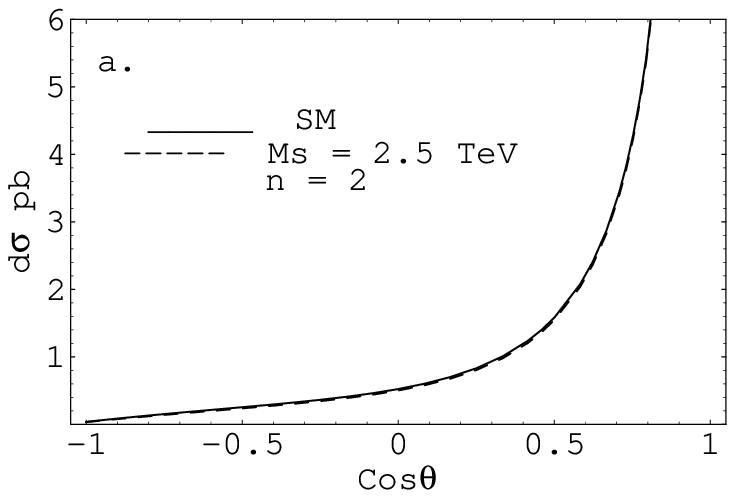}
\includegraphics{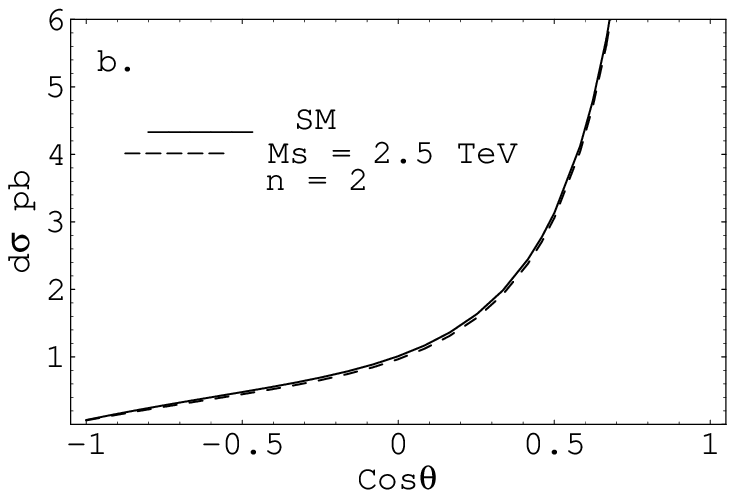}
\includegraphics{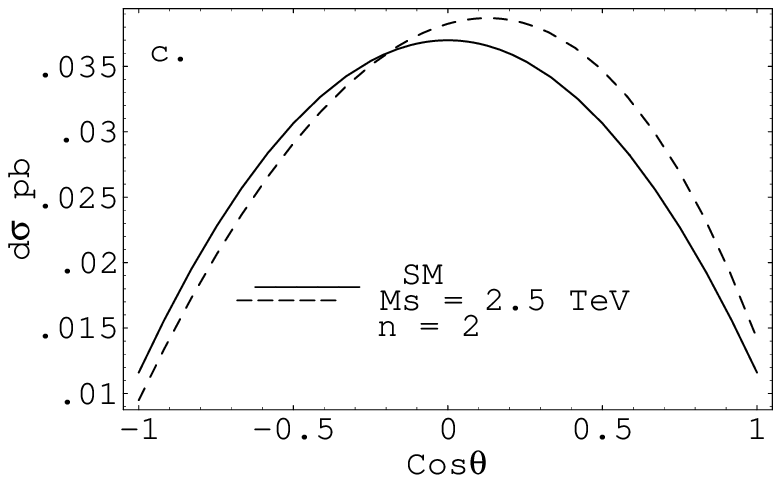}
\caption{Angular distribution of $W$'s at an LC with $\sqrt{s}=500$ GeV.
Solid curves correspond to SM, and dashed curves correspond to 
SM+Gravity with $M_S=2.5$ TeV and $n=2$.
$(a)$, $(b)$ and
$(c)$ correspond to unpolarised, left-polarised and right-polarised 
electron beams respectively. Positron beam is considered unpolarised. }
\label{fig-ang}
\end{figure}

Next we consider angular distribution of the $W$'s. Fig.\ref{fig-ang}
shows angular distributions of $W$'s for a typical value of $M_S=2$ TeV.
In the SM case there is a forward-backward asymmetry when unpolarised or 
left-polarised electron beam is considered. This is due to the $\nu$
exchange contribution.  With right-handed electron polarisation there is no
$\nu$ exchange contirbution and hence the SM distribution is symmetric. 
When graviton exchange contribution is included, the intereference of $J=2$
and $J=1$ gives rise to an asymmetric distribution even when the electron
beam is right-polarised.
This was also noticed by the authors of \cite{agashe}. 
In the following we quantify this forward-backward asymmetry.

\begin{figure}[ht]
\vskip 6cm
\includegraphics{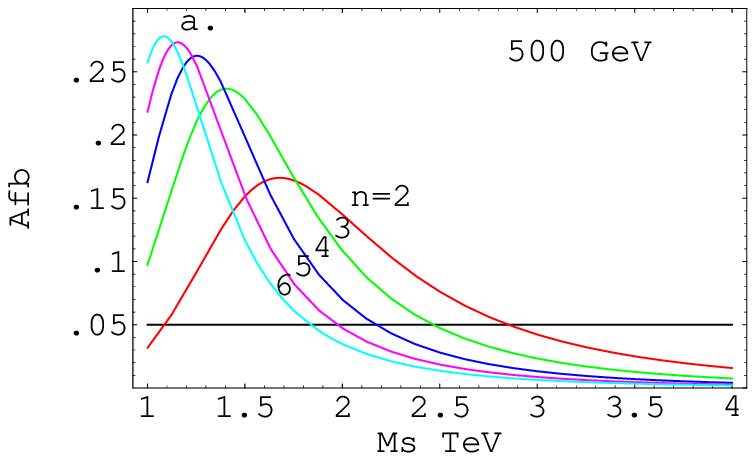}
\includegraphics{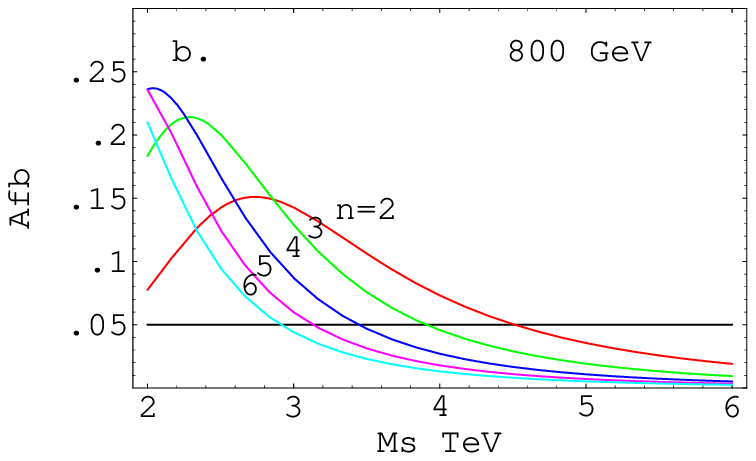}
\caption{Forward-backward asymmetry with right-polarised electron beam.
(a) $\sqrt{s}=500$ GeV  and (b) $\sqrt{s}=800$ GeV. $n$ is the number of
extra dimensions.}
\label{fig-fbasy}
\end{figure}

The forward-backward asymmetry as a function of $M_S$ is plotted in Fig.
\ref{fig-fbasy}. With $M_S=2.5$ TeV and $n=2$ asymmetry can be as large 
as 8\% at 500 GeV. The asymmetry tends to vanish as $\sqrt{s}$ aproaches 
$M_S$. This is because, then the purely gravity mediated contribution 
dominates over the interference of the SM and gravity mediated 
contribution. With $n=2$ maximum asymmetry is around 15 \% corresponding to
$M_S$ value of 1.7 (2.7) TeV at $\sqrt{s}=500$ (800) GeV.    
For the asymmetry to be at the level of 5\% or more, $M_S$ values should 
be less than 2.8 (1.8) TeV for $n=2$ (6) at a 500 GeV LC, and 
4.5 (2.9) TeV at an 800 GeV LC.

\subsection{Polarised $W$'s}

We now consider production of polarised $W$'s. $W$ polarisation can be 
measured by studying its decay distributions. We study the 
polarisation fractions, $f_{0,\pm}$ of the $W$'s produced. 
They enter in the expression for angular distribution of the secondary
lepton in the rest frame of the $W$ in the following way.

\begin{equation}
\frac{1}{\sigma} \frac{d\sigma}{d\cos\theta^*}=
\frac{3}{4}f_0\sin^2\theta^*+\frac{3}{8}f_+(1-\cos\theta^*)^2+
\frac{3}{8}f_-(1+\cos\theta^*)^2,
\label{eqn:fs}
\end{equation}

Here $\theta^*$ is the polar angle of the lepton in the rest frame of the
$W$ with $z$ axis along the boost direction.  $f_0$ gives
the fractional cross section of the longitudinal $W^-$, while $f_\pm$ give
that of the right-/left-circularly polarised $W^-$'s.
LEP2 has been able to measure the longitudinal fraction,
$f_0$ with an accuracy of 5\% \cite{LEPf0}. 

\begin{figure}[ht]
\vskip 8cm
\includegraphics{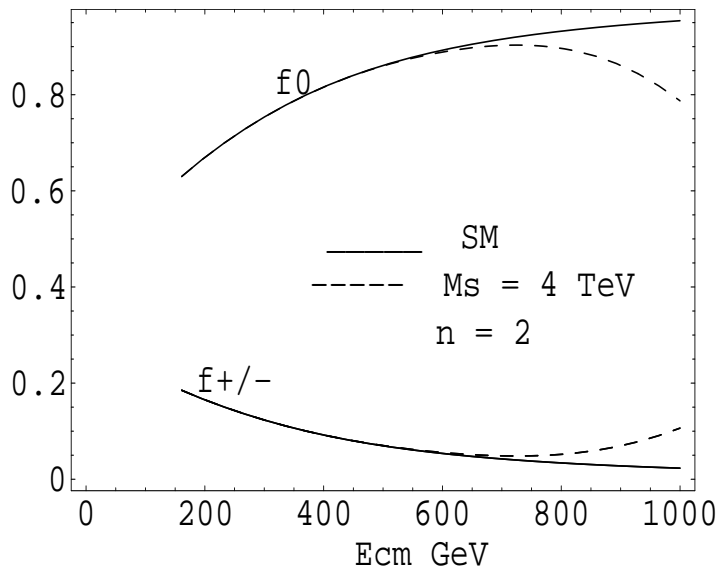}
\caption{
Polarisation fractions of $W^-$'s with right-handed electron beam.
Solid curves correspond to SM values, while dashed curves correspond 
to values with graviton exchange with $M_S=4$ TeV and number of 
extra dimensions, $n=2$ included.
}
\label{fig-fs}
\end{figure}

Table \ref{table-fs} gives 
different polarisation fractions without and with contribution due
the graviton-KK exchange at an LC with right-handed electron beam.

\begin{table}[ht]
\begin{center}
\begin{tabular}{c|c|c|c|c|c}
\hline
&&&&\\
$\sqrt{s}$&$M_S$& $f_0^{SM}$&$f_0^{G}$&$f_\pm^{SM}$&$f_\pm^{G}$ \\[2mm] 
\cline{1-6}
&&&&\\
500 GeV&2.5 TeV &0.860&0.834&0.070&0.083\\[2mm] \cline{1-6}
&&&&\\
800 GeV&4 TeV&0.932&0.897&0.034&0.052\\ [2mm] \cline{1-6}
\end{tabular}
\caption{Fractional polarisations of the $W$'s with right-handed 
electron beam. $f_{0,\pm}^{SM}$'s are SM values, 
while $f_{0,\pm}^{G}$'s those with graviton exchange included. Number 
of extra dimensions is taken to be $n=2$. Note that $f_+=f_-$.}
\label{table-fs}
\end{center}
\end{table}

Although the cross section with right-handed electron polarisation is only
about 57.1 (18.7) fb at $\sqrt{s}=500$ (800) GeV, a collider with 
luminosity of the order of 100 fb$^{-1}$ will produce several 
thousand events. 
This would allow a determination of $f_0$ with an error of 1-2 \%, 
sufficient to probe the effects of grativon exchange indicated in 
Table \ref{table-fs}.

\section{Conclusion}

We have considered the process $\eeww$ to study the effect of 
a typical low energy gravity model suggested by Arkani-Hamed, Dimopoulos 
and Dvali \cite{add}. In such a model, in addition to the usual SM
contribution, there is a contribution due to the spin-2 KK exchange in the 
$s$-channel. Earlier studies have obtained limits on the mass scale $M_S$,
at which gravity becomes comparable in strength to the other interactions,
in $4+n$ dimensions \cite{agashe,balazs}. These limits are obtained
considering deviation of the total cross section from its SM value.
In this paper we have considered observables which help distinguish the 
spin-2 nature of the interaction. A forward-backward asymmetry in the 
case of right-polarised electron beam is one significant observable in this
regard. We have quantified the asymmetry, and considered reach of $M_S$
by looking at this asymmetry at a linear collider running at $\sqrt{s}=$
500 GeV and 800 GeV. We also looked at the polarisation fractions of the 
$W$'s produced. Here again, we reach the conclusion that right-polarised
electron beam is advantageous. Our analysis shows that an $M_S$ of about
2.5 TeV could produce detectable effects at a linear collider running at 
500 GeV centre of mass energy. 
This is improved to about 4 TeV for a centre of mass energy of 800 GeV.
The effects are qualitatively different from other new physics scenarios,
{\it e.g.,} strong $WW$ interactions \cite{poulose} or extra gauge boson
models \cite{extraZ}, since these affect the $J=1$ partial waves, while
graviton exchange is specifically a $J=2$ phenomenon.

\vskip 1cm
\noindent
{\large \bf Acknowledgement}\\

I would like to thank Prof. L.M.Sehgal for suggesting this problem, and 
for many discussions.
I also wish to thank the Humboldt Foundation for a
Post-doctoral Fellowship, and the
Institute of Theoretical Physics E, RWTH Aachen for the
hospitality provided during this work.

\end{document}